\documentclass[3p,twocolumn]{elsarticle}

\usepackage{bm}
\usepackage{graphicx}
\usepackage{epstopdf}
\usepackage{amsmath}
\usepackage{amssymb}
\usepackage{xcolor}

\newcommand{\BEQ}{\begin{equation}}
\newcommand{\EEQ}{\end{equation}}
\newcommand{\BEA}{\begin{eqnarray}}
\newcommand{\EEA}{\end{eqnarray}}

\makeatletter
\newcommand\figcaption{\def\@captype{figure}\caption}

\makeatother

\begin{document}

\title{A simple spin model for three steps relaxation and secondary
  proccesses in glass formers} 
\author[sap,isc]{Andrea Crisanti}
\ead{andrea.crisanti@uniroma1.it} \author[ipcf,sap]{Luca Leuzzi}
\ead{luca.leuzzi@cnr.it} \address[sap]{Dipartimento di Fisica,
  Universit\`a di Roma ``Sapienza,'' Piazzale A. Moro 2, I-00185,
  Roma, Italy} \address[isc]{ISC-CNR, UOS {\em Sapienza}, Piazzale
  A. Moro 2, I-00185, Roma, Italy } \address[ipcf]{IPCF-CNR, UOS {\em
    Kerberos} Roma, Piazzale A. Moro 2, I-00185, Roma, Italy }

 \begin{abstract}
A number of general trends are known to occur in systems displaying
secondary processes in glasses and glass formers.  Universal features
can be identified as components of large and small cooperativeness
whose competition leads to excess wings or apart peaks in the
susceptibility spectrum. To the aim of understanding such rich and
complex phenomenology we analyze the behavior of a model combining two
apart glassy components with a tunable different cooperativeness. The
model salient feature is, indeed, based on the competition of the
energetic contribution of groups of dynamically relevant variables,
e.g., density fluctuations, interacting in small and large sets.  We
investigate how the model is able to reproduce the secondary processes
physics without further ad hoc ingredients, displaying known trends
and properties under cooling or pressing.
 \end{abstract}

\maketitle

\section{Introduction}

In several supercooled liquids and glasses processes are observed
whose typical timescales are much longer than {\em cage rattling}
microscopic motion and local rearrangement timescales of the so-called
(fast) $\beta$ processes and, yet are much shorter than the time-scale
of structural relaxation, i.e., of the $\alpha$ processes. These are
usually termed ``Secondary processes" and are related to complicated
though local (non-cooperative or not fully cooperative) dynamics. We
will, in particular, investigate Johari-Goldstein processes
\cite{Johari71}, for which special properties hold, such as 
dependence of their relaxation time on density and temperature
and a strict relationship to
 structural processes \cite{Ngai04}.
 In the present paper we will term them simply as
$\beta$, referring to {\em fast} $\beta$ processes as $\gamma$. Their
existence was first pointed out in the 1960's from dielectric loss
spectra measurements, in which they are identified by the occurrence
of a second peak at a frequency higher than the frequency $\nu_\alpha$
of the $\alpha$ processes peak.  This so-called $\beta$-peak has been
recorded in a large number of substances as, e.g., poly-alcohols
\cite{Doss02,Lunkenheimer02,Kastner11} mixtures of rigid polar
molecules and oligomers 
\cite{Blochowicz04, Blochowicz11,Prevosto07,Capaccioli11}, 
propylene glycols \cite{Koehler10} and many others
comprehensively gathered in Ref. \cite{Ngai11}.

Also in cases where the spectral density of response losses do not
clearly show a second peak, secondary processes are nkown to be active
and their presence is, then, identified, by some anomaly at frequency
higher than $\nu_\alpha$, called {\em excess wing} 
\cite{Brand99,
  Lunkenheimer02b}.  Though it was initially observed as an apart
phenomenon \cite{Wong74}, more recent investigation has shown that the
excess wings rather are manifestations of $\beta$ JG processes
\cite{Ngai03,Prevosto07, Ngai11}.  Properly tuning external parameters
(temperature, pressure, concentration, ...)  $\beta$-peaks can come
out of the excess wings or, viceversa, secondary peaks can reduce to
excess wings.  According to Cummins \cite{Cummins05} the relevant
parameter to tune in passing from one scenario to the other one might
be the rotation-translation coupling constant, becoming stronger as
density increases, and being larger for a liquid glass former made of
elongated and strongly anisotropic molecules. 

Theoretical attempts have been carried out in this direction in the
framework of Mode Coupling Theory (MCT). According to this theory the
relaxation of reorientational correlation and rotation-translation
coupling in liquids composed of strongly anisotropic molecules appears
to be logarithmic in time \cite{Goetze04}.  A comprehensive picture
is, though, not yet established and many questions are open. For
instance, about the dependence of the characteristic time-scales of JG
processes on temperature and pressure, else, on concentration.  Or
about the chance that secondary processes might disclose a certain
degree of cooperativeness \cite{Stevenson10}, or the explanation for
the persistence of the $\beta$ processes also below the calorimetric
glass transition temperature $T_g$. A very interesting question is
whether there is a straightforward connection, and, in case, which
one, between processes evolving at qualitatively different
time-scales. Were it the case, one might devise the long-time behavior
of $\alpha$ relaxation from the behaviors of the fast small-amplitude
cage dynamics ($\gamma$ processes) and of the $\beta$ secondary
processes.  In glasses, and glass formers, where $\alpha$ and $\beta$
peaks of the loss spectra can be clearly resolved in frequency one can
resort to a description based on two time-scale bifurcation
accelerations as temperature is lowered.  Processes consequently
evolve on three Òwell-separatedÓ time sectors. Examples of well
resolved peak separation can be found, e.g., in are 4-polybutadiene,
toluene \cite{Wiedersich99} , sorbitol \cite{Kastner11} mixture of
quinaldine in tri-styrene \cite{Blochowicz04, Prevosto07,
  Blochowicz11} or trimer propilene glycol \cite{Koehler10}.

A way to reproduce secondary processes, or at least some stretching in
the high frequency side of the $\alpha$ relaxation, is to include the
coupling of correlators of two different components, such has the
density correlators of tagged particles and their surrounding medium
\cite{Sjogren86}. In the limit of strong coupling between correlators
it is possible to to yield a Cole-Cole law for the loss spectrum in
the limit \cite{Goetze89a}, but no distinct apart secondary $\beta$
peak is resolved. To the aim of overcoming these limitations in the
theoretical description of secondary processes we propose a model with
a single component but a dynamic kernel corresponding to two different
kinds of cooperativeness.

\section{The model}
The model we shall discuss is known as the Spherical $s+p$ Spin Glass model 
defined by the Hamiltonian

\begin{eqnarray}
\label{f:Ham}
{\cal H} =&& -\! \sum_{i_1<\ldots <i_s}\!J^{(s)}_{i_1\ldots i_{s}}
\sigma_{i_1}\cdots\sigma_{i_s}
\\
&&
       -\!\sum_{i_1<\ldots <i_p}\!J^{(p)}_{i_1\ldots i_p}
       \sigma_{i_1}\cdots\sigma_{i_p}
       \nonumber
\end{eqnarray}
\noindent
where $ J^{(t)}_{i_1\ldots i_{t}}$ ($t=s,p$ with $s<p$ for convention)
are uncorrelated, zero mean, Gaussian variables of variance
\begin{equation}
{\overline{\left(J^{(t)}_{i_1\ldots i_{t}}\right)^2}}=\frac{J_t^2 t!}{2N^{t-1}}
\end{equation}
where the overbar denotes
the average over the quenched disorder
 and $\sigma_i$ are $N$ continuous real variables (spins) ranging from
 $-\infty$ to $+\infty$ obeying the global constraint $\sum_i
 \sigma_i^2 = N$ (spherical constraint).  The model, defined on a
 complete graph, is intrinsically mean-field.  Indeed, each
 spin interacts with all others and no geometric nor dimensional
 structure is relevant for the interaction network.  In order to
 guarantee thermodynamic convergence and an extensive energy the
 interaction magnitude is very small, and scales with the system size
 as $J_{i_1,\ldots , i_t}\sim 1/{N}^{(t-1)/2}$.

\subsection{A bit of thermodynamics}

Due to the mean-field nature of the model the metastable glassy states
responsible for the dynamic arrest can be studied by means of
thermodynamics.  Indeed, in these spherical spin models with quenched
disordered couplings, the configurational entropy, related to the
number of metastable states, is a true, static, thermodynamic state
function, unlike realistic structural glasses \cite{Leuzzi07b}.
Therefore, to make connection with glass formers, we first recall some
results on the model static properties, both in its {\em ideal} glassy
phase and in the supercooled liquid phase.  Let us define the overlap
\begin{equation}
q_{\alpha \beta}\equiv \frac{1}{N}\sum_{i=1}^N \langle\sigma_i\rangle_{\alpha}
\langle
\sigma_i\rangle_\beta
\end{equation} 
between any two glassy stable or metastable states $\alpha$ and $\beta$
whose equilibrium measure in the corresponding ergodic component is labeled by
 $\langle \ldots \rangle_{\alpha,\beta}$.

In a cooling procedure, these states first occur as excited metastable
states at the temperature $T=T_{\rm d}=T_{\rm mct}$ coinciding with
the dynamic or mode coupling temperature.  Physically, this is the
temperature at which the glassy states dominate the free energy
landscape through which the system dynamics takes place.  At this
temperature their number becomes macroscopic, i.e., exponentially
large woth the system size, and the configurational entropy (also
called complexity) becomes extensive with the system size $N$.

Below $T=T_{\rm d}=T_{\rm mct}$ the phase space breaks down into
several regions ({\sl glass} phase), and the overlap $q_{\alpha
  \beta}$ takes different values $q_\kappa$, with probability
$p_\kappa$.  The number of different values depends on both the region
of the phase diagram and the values of $s$ and $p$, and can be finite
or infinite.  In the first case the phase is called $R$ Replica
Symmetry Broken ($R$RSB), where $R$ is the number of different values
of $q_{\alpha \beta}$, while in the second case it is termed Full
Replica Symmetry Broken (FRSB). Mixed phases are also possible
\cite{Crisanti04b,Crisanti06}.

Here, we focus on the cases where secondary processes show up. As
it will be later clarified, these correspond to a static description
 in which
 $q_{\alpha   \beta}$ takes two non-trivial value ($R=2$) with probability:
\begin{equation}
\label{eq:Pq}
\begin{split}
      P(q)  = p_1\delta(q) & +(p_2-p_1)\, \delta(q-q_1)\\
                                          & +(1-p_2)\, \delta(q-q_2).
\end{split}
\end{equation}

\begin{figure}[t!]
\includegraphics[width=.99\columnwidth]{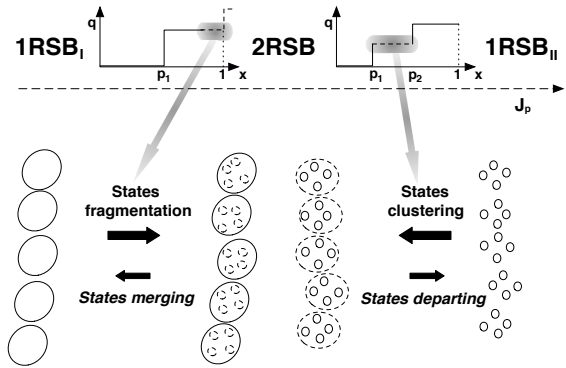}
\caption{Pictorial representation of the inverse $q(x)$ of the cumulative
  function $x(q)$ of the $P(q)$ defined in Eq. (\ref{eq:Pq}) as $J_p$ is
  increased keeping $J_s$ fixed. Below, the inner and outer states in
  the 2RSB case are shown along with their ``generation'' by fragmentation or
  clustering moving from 1RSB sectors of the phase diagram, where no
  secondary processes occur. }
\label{fig:frag}
\end{figure}

In terms of free energy lanscape, this picture displays a hierarchical
structure where each glassy minimum - representing a separated ergodic
component - also contains a further set of glassy minima inside, as
pictorially represented in Fig. \ref{fig:frag},
cf. Ref. \cite{Leuzzi08b}.  Inner states have higher overlap $q_2$
while outer states have lower overlap, $q_1<q_2$.  Such ``nested''
minima appear as metastable states at the tricritical point along the
dynamic (swallowtail) arrest line. They become the {\em ideal
  equilibrium} stable glass states at the static - Kauzmann -
transition line.  As the dynamic arrest line is approached, e.g., by
cooling, far from the tricritical point only one set of states appear.
This implies that only one kind of diverging timescale occurs for slow
processes, the structural ones.

The values of  $q_1$ and $q_2$  are given by the solution of the 
self-consistent equations  \cite{Crisanti07b,Leuzzi08b}:
 \begin{eqnarray}
 &&\hspace*{-1cm}{\cal M}(q_1)=\frac{q_2-q_1}{\chi(q_1)\, \chi(0)},\\
  &&\hspace*{-1cm}{\cal M}(q_2)-{\cal M}(q_1)=\frac{q_2-q_1}{\chi(q_2)\,
\chi(q_1)},
 \end{eqnarray}
where 
\begin{eqnarray}
\chi(q_2)&=&1-q_2,
\nonumber\\
\chi(q_1)&=&\chi(q_2)+p_2(q_2-q_1),
\nonumber
\\
\chi(0)&=&\chi(q_1)+p_1q_1,
\nonumber
\end{eqnarray}
and where we have introduced the functions
\begin{eqnarray}
&&\hspace*{-.76cm}g(X)=\frac{v_{s-1}}{s}X^{s}+\frac{v_{p-1}}{p}X^{p},
\\
&&\hspace*{-1cm}{\cal M}(X)= g'(X)=v_{s-1} X^{s-1}+v_{p-1}X^{p-1}
\label{f:emme}
\end{eqnarray}
with 
\begin{eqnarray}
v_{t-1}&=&t\beta^2 J_t^2/2.
\label{f:mct_param}
\end{eqnarray}
In a pure static study, the thermodynamics is ruled by the states that
extremize the free energy, and this leads to the two additional
self-consistent equations ({\sl static} condition)
 \begin{eqnarray}
&&\hspace*{-1cm}g(q_2)-g(q_1)=(q_2-q_1)\left[{\cal M}(q_1)-\frac{1}{p_2\chi(q_1)}\right]
\nonumber
\\
&&\hspace*{1cm}-\frac{1}{p_2^2}\ln
\frac{\chi(q_2)}{\chi(q_1)},
\label{f:stat1}
 \\
 &&\hspace*{-1cm}g(q_1)=-\frac{q_1}{p_1\chi(0)} 
-\frac{1}{p_1^2}\ln
\frac{\chi(q_1)}{\chi(0)},
\label{f:stat2}
 \end{eqnarray}
which fix the value of $p_1$ and $p_2$. Non-trivial solutions of these
equations appear at the static (else called  Kauzmann) critical temperature
$T=T_K < T_{\rm d}$.

To account for the metastable states that dominate the dynamics in
the $2$RSB phase for temperatures $T\in[T_K,T_{\rm d}]$, equations
(\ref{f:stat1},\ref{f:stat2}) must be replaced by
\begin{eqnarray}
{\cal M}'(q_\kappa)=\frac{1}{(1-q_\kappa)^2}, \quad \kappa=1,  2
\end{eqnarray}
which follows from the requirement that the solution {\sl maximizes}
the complexity \cite{Crisanti03b}.  This condition ensures that the
dynamics ruled by the memory kernel ${\cal M}(\phi)$ be marginally
stable \cite{Crisanti07b}. It is known as the {\sl marginal condition}
because it coincides with the marginal stability condition in the
solution of the statics of the model \cite{Crisanti07a}.  It bridges
the static and dynamic properties in the $2$RSB phase, where $q_1$ and
$q_2$ become the two nontrivial asymptotic {\em plateau} values,
i. e., non-ergodicity factors, of the dynamic correlation function in
the three steps relaxation scenario.

Away from the $2$RSB phase, only one plateau occurs. In these cases
the solutions to the above equations coincide $q_1=q_2$.  Beyond the
dynamic critical line glass-to-glass transitions can occur, between
$1$RSB and $2$RSB kind of glasses. Here we are mainly interested in
the equilibrium dynamics of supercolled liquids. The interested reader in the 
frozen glass phase can look, e.g.,  at Refs. \cite{Crisanti06,Crisanti07b}.

\subsection{Dynamic phase diagrams and swallowtail singularity}

The existence of two nontrivial asymptotic plateau's of the dynamic
correlation function, approaching the $2$RSB phase, is associated with
the presence of a given type of singularity in the dynamic equations.
According to Arnold's classification of singular points in
catastrophe theory, the model has to display a double bifurcation $A_4$,
or {\em swallowtail}, singularity.

A static glass $2$RSB phase in the spherical $s+p$ spin glass model
can, actually, be found provided $s$ and $p$ are equal or larger than the
solution of
\begin{equation}
(p^2+s^2+p+s-3ps)^2-ps(p-2)(s-2)=0,
\end{equation}
as it has been shown in Ref. \cite{Crisanti07b}.  Some {\sl threshold}
values of $(s,p)$ are $(3,8)$, $(4,11)$ or $(5,16)$.  The larger
$p-s$, the broader the region of phase diagram where the static $2$RSB
phase can be found. This is a necessary condition for the occurrence
of a $2$RSB phase {\em somewhere} in the static (thermodynamic) phase
diagram but do not guarantee the occurrence of an $A_4$ singularity
along the dynamic arrest line.

To have a $2$RSB phase accessible in the MCT equilibrium dynamics -
i.e., a swallowtail singularity along the dynamic arrest line - the
condition on $s$ and $p$ is stronger \cite{Krakoviack07b}:
\begin{equation}
\sqrt{(s-1)(p-1)}-\sqrt{(p-2)(s-2)}\geq \sqrt{2}.
\end{equation}
 In this case a the $A_4$ point is exposed to the fluid phase and a
 three step correlation function, or three peak loss function, develops
 approaching the dynamic transition next to this point. Some 
lower bound values are $(s,p)=(3,10)$, $(4,16)$, $(5,22)$.

Moreover, in order to have a swallowtail also in the static-Kauzmann
transition line the parameters $s$ and $p$ must further satisfy the
equation
\begin{equation}
  (sp - p - s +1) y^2 - (p+s+1) y + 2 \geq 0
\end{equation}
where $y \in [0,1]$ is solution of 
\begin{equation}
 (sp - p - s -1) y + p + s - 1 = sp\, z(y)
\end{equation}
and $z(y)$ is the CS $z$-function \cite{Crisanti92}
\begin{equation}
  z(y) = -2y \frac{1-y+\ln y}{(1-y)^2}.
\end{equation}
Some critical values $(s,p)$: $(3, 13)$, $(4, 23)$, $(5,35)$.  In this
case the stable {\em ideal} 2RSB phase can be accessed directly from
the stable fluid phase.

Relevant external parameters  for the phase diagram will be the
``concentration" of large cooperativeness $\rho$, defined as
\begin{eqnarray}
J_p &=&\rho J\quad;\quad J_s = (1-\rho) J\\
J&=&J_s+J_p 
 \end{eqnarray}
 and the temperature, defined in units of the small cooperativeness
 interaction $J_s$,
 \begin{eqnarray}
 \frac{T}{J_s} =\sqrt{\frac{s}{2v_{s-1}}} 
 \label{f:temp}
\end{eqnarray}

\begin{figure}[t!]
\includegraphics[width=.99\columnwidth]{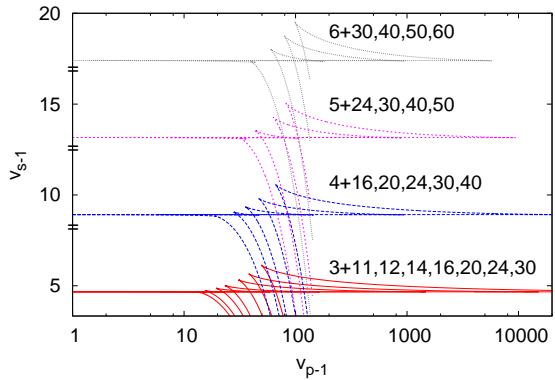}
\caption{Dynamic phase diagrams in the $v_s,v_{p}$ plane.  The dynamic
  lines with swallow tail are represented for series of models with
  $s=3, 4, 5$ and $6$.}
\label{fig:vs_vp}
\end{figure}

\begin{figure}[t!]
\includegraphics[width=.99\columnwidth]{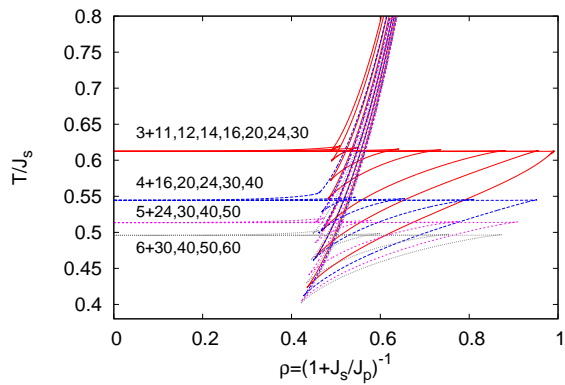}
\caption{Dynamic phase diagrams in the $T/J_{s},\rho$ plane for different sets of models.}
\label{fig:T_rho}
\end{figure}

Transition lines can be drawn parametrically in the overlap variable
$q\in[0,1]$ at the dynamic arrest fold singularity,
cf. Fig. \ref{fig:vs_vp}: \cite{Crisanti07c}
\begin{eqnarray}
v_{p-1}&=&\frac{(s-1)q-(s-2)}{(p-s)q^p-2(1-q)^2}
\\
v_{s-1}&=&\frac{(p-1)q-(p-2)}{(p-s)q^s-2(1-q)^2}
\end{eqnarray}
or in $T, \rho$ using the transformations Eqs. (\ref{f:mct_param}),
(\ref{f:temp}), as shown in Fig. \ref{fig:T_rho}.


\section{Dynamics equations}
To study the slowing down of the dynamics as the critical arrest
is approached from the liquid phase, we cannot rely only on the static
analysis and the dynamics of the model must be analyzed.  The
relaxation dynamics of the system is described by the Langevin
equation
\begin{eqnarray}
&&\Gamma_0^{-1}\frac{\partial\sigma_k(t)}{\partial t}= -\frac{\delta {\cal H}[\{\sigma\}]}{\delta \sigma_k(t)}+\eta_k(t)
\\
\nonumber
&&\qquad\langle \eta_k(t)\eta_n(t')\rangle=2 k_B T \Gamma_0^{-1} \delta_{kn}\delta(t-t')
\end{eqnarray}
where $\eta_k$ is the thermal white noise and $\Gamma_0^{-1}$ is the
microscopic time-scale.  Using the Martin-Siggia-Rose response field
approach in the path-integral formalism \cite{Martin73,Dedominicis78},
the average over the quenched disorder can be performed, and the
equations of motion reduce to the self-consistent dynamics of single
variable $\sigma(t)$.  The fundamental observables to study the onset
of the dynamic slowing down are the diagonal spin-spin time
correlation function $C(t,t')$ and the spin-response function
$G(t,t')$, which for our model are defined as\footnote{We have
  included the temperature into the definition of the response
  function.}
\begin{eqnarray}
C(t,t') &=& {\overline {\langle \sigma(t)\sigma(t') \rangle}},
\\
G(t,t')&= & \left. \frac{\delta\overline{\langle \sigma(t)\rangle}}{\delta \beta h(t')}\right|_{h=0},
 \qquad t>t',
\end{eqnarray}
with $C(t,t) = 1$ from the spherical constraint.  The brackets denote
the thermal average over different trajectories (and initial
conditions).  For temperature above $T_{\rm d}$ the dynamics is time
translational invariant (TTI) and the response and correlation
functions are related by the Fluctuation - Dissipation Theorem (FDT):
\begin{equation}
G(t-t')=\theta(t-t')\frac{\partial C(t-t')}{\partial t'}.
\end{equation}
In this case, and using the shorthands $F'(t)\equiv \partial F(t)
/\partial t$, the dynamic equation for $C(t-t')$ takes the form
\begin{eqnarray}
&&\hspace*{-1cm}\Gamma_0^{-1} \partial_t C(t)+\bar r C(t)+\int_0^t d t'{\cal M}[C(t-t')]\, C'(t')
\nonumber
\\&&\hspace*{4cm}= \bar r -1
\label{eq:dyn}
\end{eqnarray}
with initial the condition $C(t=0)=1$ and 
\begin{equation}
\bar r = r-{\cal M}[C(t=0)]
\label{eq:bar_r}
\end{equation}
The parameter $r$ in the above equation 
is a ``bare mass" \cite{Crisanti08} related to the
Lagrange multiplier needed to impose the spherical constraint
\cite{Crisanti93}. The value of 
$\bar r$ can depend on temperature and on
$\rho$ through ${\cal M}[C(t=0)]$. However, 
above $T_d$, $\bar r$ is constant and equal to $1$, so
that the r.h.s. of (\ref{eq:dyn}) vanishes.

The kernel memory function ${\cal M}(t)\equiv {\cal M}[C(t)]$ for the
spherical $s+p$ spin glass model we are considering has the functional
form shown in Eq. (\ref{f:emme}).  We stress that eq. (\ref{eq:dyn})
(with $\bar r=1$) is the equation governing the time correlation
function in a schematic mode-coupling theory in which the second order
time derivative term of the Mode Couplings equations is replaced by a
first order one \cite{Bouchaud96,Goetze09}.

To discuss the slowing down of the dynamics as the critical point is
approached it is, further,  useful to introduce the function $\bar r(q)$
\cite{Crisanti92}
\begin{equation}
\bar r(q) \equiv \frac{1}{1-q}-{\cal M}(q),
\end{equation}
which determines the asymptotic value of the correlation function.
Indeed,
it can be shown that in the long time limit the asymptotic value  
$q=\lim_{t\to\infty}C(t)$ of the correlator $C(t)$ solution of Eq. (\ref{eq:dyn})
is given by the condition:
\begin{equation}
\label{eq:rbar}
\bar r(q)=\bar r.
\end{equation}
where $\bar r$ is the parameter appearing in Eq. (\ref{eq:dyn}) and
defined in Eq. (\ref{eq:bar_r}).  The additional requirement for a
critical dynamics, the marginal condition, imposes that $\bar r(q)$ be
a local minimum with $\bar r^\prime(q)=0$ for $q$ solution of
Eq. (\ref{eq:rbar}).  When this happens $C(t)$ develops a plateau at
$C(t)=q$, or multiple plateaus if more solutions to
Eq. (\ref{eq:rbar}) become marginal simultaneously, as it is the case
of the $2$RSB phase that we are discussing.


The properties of the dynamics close to the critical point can be analyzed by
writing $C(t)=q+\phi(t)$, where $q$ is generic for the moment,  
and expanding ${\cal M}[C(t)]$ for small $\phi$:
\begin{equation}
{\cal M}(q+\phi)=\sum_{m=0}^\infty\frac{{\cal M}^{(m)}(q)}{m!}\phi^m
\end{equation}
where 
\begin{equation}
\label{eq:der}
\begin{split}
{\cal M}^{(m)} &=\frac{d^m {\cal M}(q)}{dq^m} \\
                       & = \frac{m !}{(1-q)^{m+1}} - \frac{d^m \bar r(q)}{dq^m}.
                       \end{split}
\end{equation}
The integral term in  Eq. (\ref{eq:dyn}) then reads 
\begin{eqnarray}
\label{f:integral}
&&\hspace*{-1cm}
\int_0^t dt' {\cal M}[C(t-t')] C'(t')=
\nonumber
\\
\nonumber
&&\hspace*{-.5cm}\sum_{m=1}^\infty \left[
\frac{{\cal M}^{m-1}(q)}{(m-1)!}-(1-q)\frac{{\cal M}^{m}(q)}{m!}
\right]\phi^m(t)
\\
&&\hspace*{-.5cm}-(1-q){\cal M}(q)  
+\sum_{m=1}^\infty\frac{{\cal M}^{m}(q)}{m!}I_m(t)
\end{eqnarray}
where
\begin{equation}
I_m(t)\equiv\int_0^t\, dt^\prime\,\left[ \phi^m(t - t^\prime) - \phi^m (t) \right]\phi^\prime(t^\prime),
\end{equation}
leading, after few algebraic manipulations, to:
\begin{eqnarray}
\label{eq:scale1}
\nonumber
&&\hspace*{-1cm}\Gamma_0^{-1}\phi'(t) + \left[ \bar r + {\cal M}(q) - (1-q) {\cal M}^{(1)} (q) \right] \phi(t)  \\ \nonumber
&&\hspace*{-1cm}\quad+ \sum_{m=2}^\infty\left[ \frac{{\cal M}^{(m-1)}}{(m-1)!} - (1-q) \frac{{\cal M}^{(m)}(q)}{m!} \right]\phi^m(t)   \\ 
&&\hspace*{-1cm}\quad + \sum_{m=1}^\infty \frac{{\cal M}^{(m)}(q)}{m!}I_m(t)=
\\
\nonumber&& \hspace*{2.5cm}(1-q)\left[ \bar r + {\cal M}(q) \right] -1\, .
\end{eqnarray}

The final step replaces the kernel ${\cal M}(t)$  in terms of $\bar r(q)$ 
by writing, see  Eq. (\ref{eq:der}),
\begin{equation}
\frac{{\cal M}^{(m)}(q)}{m!}=\frac{1}{(1-q)^3}\left( \gamma_m - \delta_m \right)\, .
\label{eq:lambda}
\end{equation}
with
\begin{eqnarray}
\gamma_m &\equiv& \frac{1}{(1-q)^{m-2}}\\ \nonumber
\delta_m &\equiv& \frac{(1-q)^3}{m!}\frac{d^m}{dq^m}\left[ \bar r(q) - r\right],
\end{eqnarray}
leading to:
\begin{eqnarray} 
&&\hspace*{-.5cm}\Gamma_0^{-1} \phi'(t) 
\\
\nonumber
&&- \frac{1}{(1-q)^3}\sum_{m=1}^\infty\left[ \delta_{m+1} -
 (1-q)\delta_m \right] \phi^m(t) \\ 
&&+\frac{1}{(1-q)^3}\sum_{m=1}\left[ \gamma_m - \delta_m \right] I_m(t) = 
\nonumber
\\
&&\hspace*{4.5cm}-\frac{\delta_0}{(1-q)^2}.
\nonumber
\end{eqnarray}

This equation describes the behavior of $C(t)$ close to a generic $q
\in [0,1]$.  In the ergodic liquid (paramagnetic) phase above $T_{\rm
  d}$ the asymptotic value of $C(t)$ is $q=0$. However, as the critical
point where the dynamical critical arrest occurs is approached, one (or
more) value of $q > 0$ appears where $\bar r(q)-\bar r \ll 1$ and
\begin{equation}
\delta_1 = \bar r^\prime(q)=0.
\end{equation}
Then, we  introduce the small parameter 
\begin{equation}
\sigma\equiv \delta_0=(1-q)^3\left[ \bar r(q)-\bar r \right]  \ll 1\, .
\end{equation}
and write
\begin{equation}
 \delta_2=1-\lambda,
\end{equation}
where 
\begin{equation}
\lambda \equiv \frac{(1-q)^3}{2}{\cal M}''(q)
\end{equation}
is called the {\sl exponent parameter}.
The dynamics close to the plateau $C(t) = q$ is, therefore, ruled by 
\begin{eqnarray} 
&&\hspace*{-1cm}\Gamma_0^{-1}\phi'(t) - \frac{\sigma}{(1-q)^3} \phi(t) 
\label{eq:scale3}
 \\ 
\nonumber
&&\hspace*{-.8cm}+\frac{1}{(1-q)^2}\left[ (1-\lambda)\phi^2 (t)+ I_1(t)\right] + O(\phi^3)=
\\
&&\nonumber \hspace*{4cm} -\frac{\sigma}{(1-q)^2}  \ .
\end{eqnarray}
For $\sigma\ll 1$ the solution of Eq. (\ref{eq:scale3}) assumes the scaling form
\begin{equation}
\label{eq:scaling}
\phi(t)= \sigma^{1/2} g(\tau)\, , \;\; \tau=t/t_\sigma=o(1)
\end{equation}
with $g(\tau)$ solution of the  scaling equation:
\begin{equation}
\nonumber
 \int_0^\tau\!\!\! d\tau' \left[ g(\tau-\tau') - g(\tau) \right]  g^\prime(\tau')+(1-\lambda) g^2(t)=-1,
\label{eq:rescaled1}
\end{equation}
and $t_\sigma$ diverging at the critical point $\sigma\to 0$.
\begin{figure}[t!]
\includegraphics[width=.99\columnwidth]{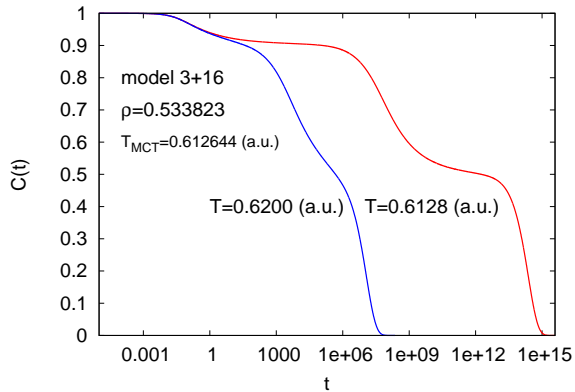}
\caption{Correlation functions for the model $3+16$: two step relaxation is evident as the temperature is lowered. The exponent parameters are  $\lambda_{2}=0.689542$, $\lambda_1=0.538201$.}
\label{fig:ct_3_16}
\end{figure}

\begin{figure}[t!]
\includegraphics[width=.99\columnwidth]{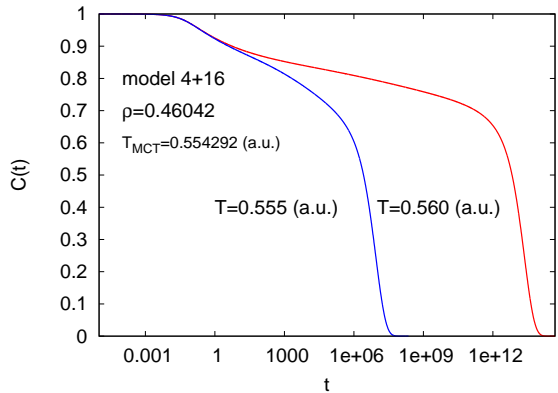}
\caption{Correlation functions for the model $4+16$: the two step
  relaxation is practically invisible because of the merging of two
  nearly logarithmic decays. Indeed, the model parameter exponents are
  $\lambda_2= 0.993307$ and $\lambda_1= 0.992734$ for this model. }
\label{fig:ct_4_16}
\end{figure}

\subsection{Three step relaxation}
Close to the transition point to a $2$RSB phase the correlation
function $C(t)$ develops two plateaus for $C(t) = q_\kappa$ (with
$\kappa=1,2$), with
\begin{equation}
 \bar r(q_1) = \bar r(q_2)\ \mbox{and}\ 
 \bar r'(q_1) = \bar r'(q_2)= 0
\end{equation}
Near each plateau $q_k$ the scaling solution (\ref{eq:scaling}) predicts 
for $C(t) \gtrsim q_k$ the power law behavior:
\begin{equation}
C(t) -  q_\kappa \sim   t^{-a_\kappa},
\label{eq:a_k}
\end{equation}
with the exponent $ 0<a_\kappa<1/2$ fixed by 
\begin{equation}
\lambda_\kappa=
\frac{\Gamma^2(1-a_\kappa)}
{\Gamma(1-2a_\kappa)},
\end{equation}
where 
\begin{equation}
\lambda_\kappa = \frac{(1-q_\kappa)^3}{2}{\cal M}''(q_\kappa).
\label{eq:exact_lambda}
\end{equation}
For $C(t) \lesssim q_\kappa$ the scaling solution leads the von
Schweidler law
\begin{equation}
C(t) - q_\kappa \sim -t^{b_\kappa}
\label{eq:b_k}
\end{equation}
with the exponent $0<b_\kappa<1$ obtained from
\begin{equation}
\lambda_\kappa=\frac{\Gamma^2(1+b_\kappa)}{\Gamma(1+2b_\kappa)}.
\end{equation}

In Figs. \ref{fig:ct_3_16}, \ref{fig:ct_4_16} we show the numerical
solution of Eq. (\ref{eq:dyn}) close to the dynamical arrest for a
system with a $2$RSB phase showing a three steps relaxation and a
system displaying a nearly logarithmic decay.

\begin{figure}[t!]
\includegraphics[width=.99\columnwidth]{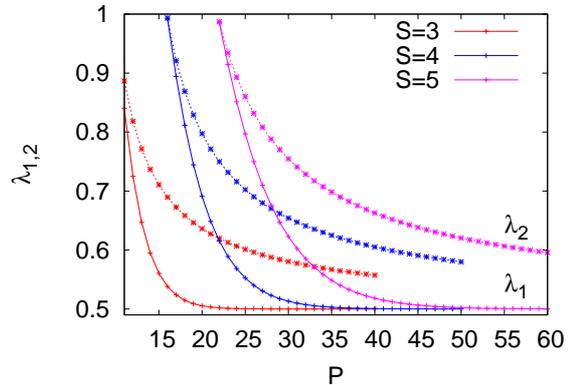}
\caption{The behavior of the exponent parameters $\lambda_1$ (bottom)
  and $\lambda_2$ (top) for different models displaying three step
  relaxation.}
\label{fig:lambdas}
\end{figure}

In Fig. \ref{fig:lambdas} we plot the $p$ dependence, at given $s$ for
the MCT parameter exponent $\lambda_{1,2}$.

\section{Susceptibility spectra}
\begin{figure}[t!]
\includegraphics[width=.99\columnwidth]{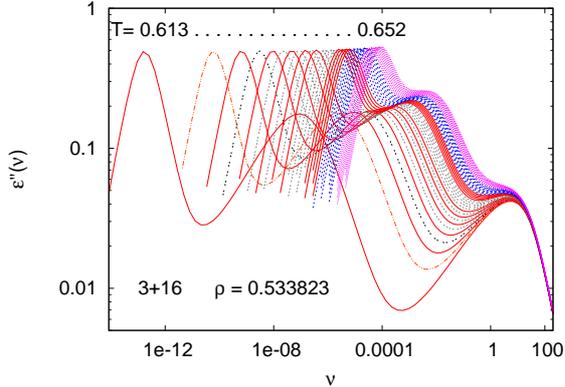}
\caption{Suceptibility loss for the $3+16$ model in a cooling
  procedure towards the $A_4$ singularity. The $\beta$ peak appears in
  between $\alpha$ and $\gamma$ peaks as $T$ is decreased.}
\label{fig:spectra_3_16}
\end{figure}

\begin{figure}[t!]
\includegraphics[width=.99\columnwidth]{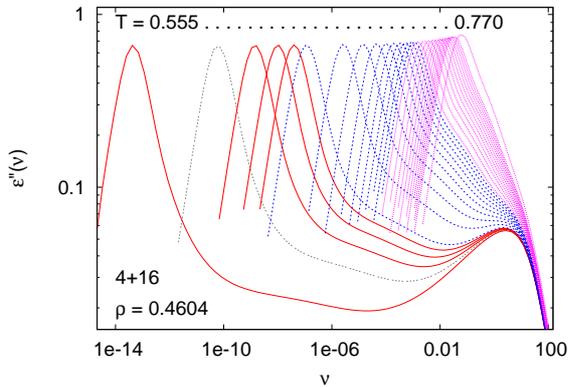}
\caption{Suceptibility loss for the $4+16$ model in a cooling procedure towards the $A_4$ singularity. No full $\beta$ peak develops in cooling. } 
\label{fig:spectra_4_16}
\end{figure}

Experimental data are more often available in the frequency domain
rather than in the time domain. We show in Figs. \ref{fig:spectra_3_16},
\ref{fig:spectra_4_16} the behavior of the susceptibility in cooling
procedures towards the $A_4$ singularity where the $\beta$ peak, if
existing, is most prominent. The two cases are qualitatively quite
different. In the case shown in Fig. \ref{fig:spectra_3_16}, the $3-16$
model, the onset of a $\beta$ peak is quite tidy.  In the $4+16$ case,
reported in Fig. \ref{fig:spectra_4_16} no $\beta$ peak is evident not
even at very low temperatures and a kind of excess wing appears in its
place.


\begin{figure}[t!]
\includegraphics[width=.99\columnwidth]{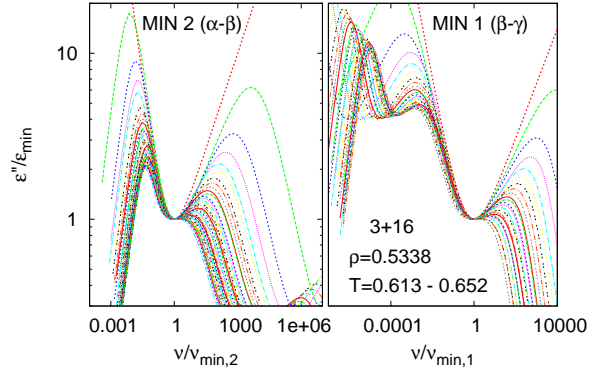}
\caption{Rescaled loss spectra next to the minima between $\alpha$ and $\beta$ peaks, left, and $\beta$ and $\gamma$ peaks, right. }
\label{fig:scalingLoss}
\end{figure}

For what concerns loss spectra, the Mode Coupling scaling next to a $A_4$
point in the $\omega$ space, that is next to the minima $\epsilon_{\rm
  min}^\kappa$ of the dynamic susceptibility $\epsilon''(\nu)$,
becomes \cite{Goetze09}:
\begin{eqnarray}
&&\hspace*{-1.2cm}\epsilon''(\nu)=\frac{\epsilon_{\rm min}^\kappa}{a_\kappa+b_\kappa}\left[
a_\kappa\left(\frac{\nu}{\nu_{\rm min}^\kappa}\right)^{-b_\kappa}
+b_\kappa \left(\frac{\nu}{\nu_{\rm min}^\kappa}\right)^{a_\kappa}
\right]
\nonumber
\\
&& \kappa=1,2
\end{eqnarray}
where the height of the minimum scales as $\epsilon_{\rm min}\propto
\sqrt{T-T_{\rm mc}}$ and the position of the frequency goes like
\begin{equation}
\nu_{\rm min}^\kappa\propto \left(T-T_{\rm mc}\right)^{1/(2a_\kappa)}
\end{equation}
In figure \ref{fig:scalingLoss} we show an instance of such scaling next
to the $A_4$ singularity point for the $3+16$ model.

\section{Multi-scale and strechted relaxation}

In the mean-field schematic MCT the liquid glass former is homogeneous. 
Different characteristic relaxation times can occur because of the interplay of 
different relaxation mechanisms taking place homogeneously in space. Indeed,
because of the mean-field nature of MCT \cite{Andreanov09}, position space 
does not play any role.

In order to have more relaxation times in MCT one has, thus, to resort
to a schematic model with a memory kernel more complicated than a
simple power of the time autocorrelation function, including at least
two parameters, cf. Eq. (\ref{f:emme}), or including
more components, that is involving the correlation of different
degrees of freedom \cite{Sjogren86}.

For instance, a $F_2$ theory \cite{Leutheusser84,Bengtzelius84}
displays dynamic arrest at a certain fold singularity $A_2$, denoted
by the mode coupling temperature $T_{\rm mc}$ but the relaxation is
Debye (a simple exponential in the time domain).

A $F_{12}$ theory \cite{Goetze84,Kirkpatrick88a} {\em colors} the
$\alpha$ relaxation to something that can be interpreted, i.e.,
numerically interpolated, as a stretched Kohlrausch-Williams-Watts
exponential. The link is provided by setting a correspondence between
the $b$ exponent of the von Schweidler law decay from the correlation
function plateau and the $\beta_{\rm KWW}$ exponent
\cite{Goetze09}. However, this only holds next to the plateau (or to a
minimum in the loss spectrum), whereas the long time correlation
eventually relaxes to zero as an exponential (or a Debye peak in the
low frequency domain).

In a two dimensional parameter space $(v_s,v_p)$ we observe that
enhancing the difference between the powers in the kernel
(\ref{f:emme}) the strechted KWW relaxation (and any related
Cole-Cole, or Cole-Davidson or Havriliak-Negami spectrum
\cite{Donth01}) is just an artifact of interpolation.

\begin{figure}[t!]
\includegraphics[width=.99\columnwidth]{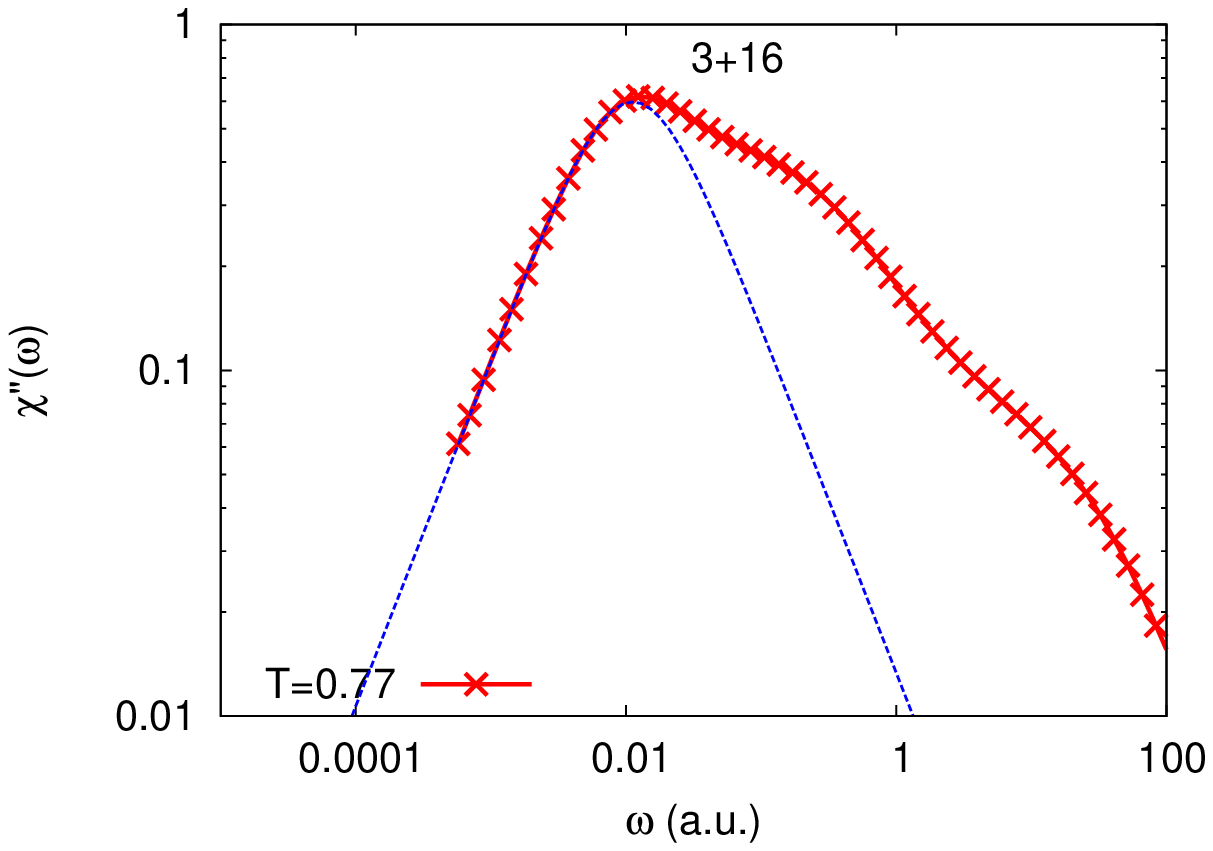}
\includegraphics[width=.99\columnwidth]{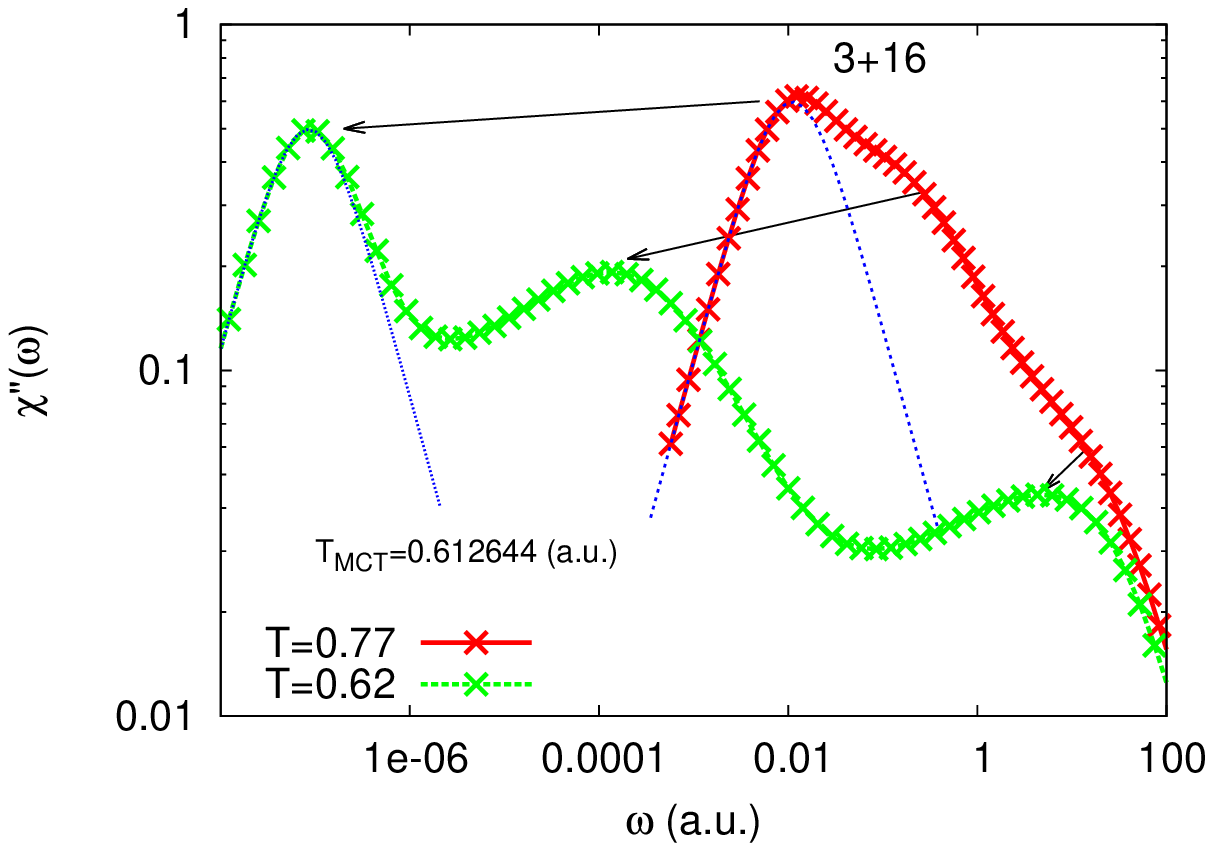}
\caption{Low frequency Debye relaxation in the $(3,16)$ model. In the
  supercooled liquid at not very low temperature the beta peak is
  hidden in the high frequency tail of the alpha relaxation (top),
  whereas, as temperature is sensitively lowered towards $T_{\rm mc}$
  the $\beta$ peak emerges. }
\label{fig:colecole}
\end{figure}

For $p\gg s$ we have a $A_4$ singularity. In the near proximity to
this point dynamic arrest can occur at two different plateaus, each
one with its critical slowing down exponents $a_{1,2}$, $b_{1,2}$, and
characteristic temperature scalings of the relaxation time
$\tau_{1,2}\propto (T-T_c)^{\gamma_{1,2}}$.  Both relaxations are well
separated in time, and in frequency, where a minimum of
${\chi(\omega)''}$ corresponds to each plateau in $\phi(t)$,
cf. Fig. \ref{fig:colecole}. The low frequency susceptibility peaks
are, however, clearly Debye. In the time domain this means that, apart
from the approach to/decay from the plateaus, the relaxation is
exponential.

\begin{figure}[t!]
\includegraphics[width=.99\columnwidth]{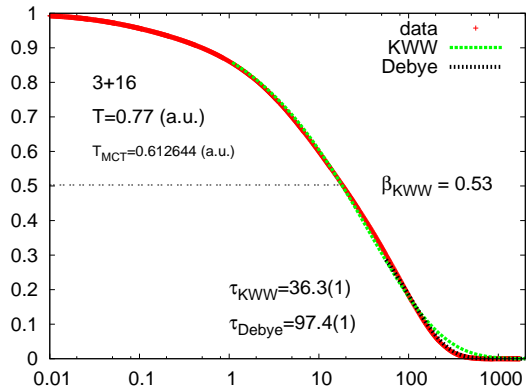}
\caption{Instance of a very good stretched exponential interpolation
  of pure two exponentials in the $(3,16)$ model.}
\label{KWW}
\end{figure}

As we measure correlations and spectra a bit further away from the
$A_4$ point, though, the two relaxations mix yielding a very well
interpolated stretched exponential as shown in Fig. \ref{KWW}.

\section{Conclusions}
In this work we have shown how some features of secondary process
observed in the relaxation of glass and glassy systems can be captured
by a simple schematic model.  The model, known as the Spherical $s+p$
Spin Glass model, is a mean field model whose static and dynamics
properties can be worked out analytically. In particular its
relaxation dynamics is described by the MCT equation with a non-linear
memory kernel sum of the a $s-1$ and $p-1$ power. Depending on the
value of $s$ and $p$ different scenarios are possible.  Secondary
processes are observed for $s$ large enough and $p\gg s$.  Here we
have focused on the properties of the model, connection with
experiments will be addressed in a future work.

Eventually we comment of the possibility of describing hierarchies of
apart secondary processes, both with discrete or continuous
time-scale separation. Though the discrimination of such phenomena is
actually rather difficult in experiments, different versions of the present 
$s+p$ models might straightforwardly account for them \cite{Crisanti07a}.

Further work on non-equilibrium dynamics and aging in $s+p$ models
with secondary processes is currently in preparation.

\bigskip

\noindent {\em Acknowledgments ---} The research leading to these results has
 received funding from the People Programme (Marie Curie Actions) of
 the European Union's Seventh Framework Programme FP7/2007-2013/ under
 REA grant agreement no. 290038 - NETADIS project, from the European Research
Council through ERC grant agreement no. 247328 - CriPherasy project, and from the Italian
 MIUR under the Basic Research Investigation Fund FIRB2008 program,
 grant No. RBFR08M3P4, and under the PRIN2010 program, grant code
 2010HXAW77-008.
 AC acknowledge financial support from European Research
Council through ERC grant agreement no. 247328 


\end{document}